\documentclass[aps,prl,showpacs,twocolumn]{revtex4}
\input epsf.sty
\def\MP{\mbox{$M_P$}}
\def\mp{\mbox{$M_P$}\ }
\def\TH{\mbox{$T_H$}\ }
\def\mbh{\mbox{$M_{\rm BH}$}\ }     
\def\MBH{\mbox{$M_{\rm BH}$}}         
\def\MET{\mbox{${\hbox{$E$\kern-0.6em\lower-.1ex\hbox{/}}}_T$}} 
\def\met{\mbox{${\hbox{$E$\kern-0.6em\lower-.1ex\hbox{/}}}_T$}\ } 
\def\ifb{fb$^{-1}$}                     

\newcommand{\AmS}{{\protect\the\textfont2
   A\kern-.1667em\lower.5ex\hbox{M}\kern-.125emS}}

\begin{document}
\title{Black Holes at the LHC
\vspace{-0.2in}}

\author{Savas Dimopoulos$^{a\dagger}$ and Greg Landsberg$^b$}\thanks{Results presented at the Les Houches Workshop ``Physics at the TeV Colliders'' (May 30, 2001) and the ``Avatars of M-Theory'' conference, ITP at Santa Barbara (June 7, 2001),
{\footnotesize http://online.itp.ucsb.edu/online/mtheory\_c01/dimopoulos}.}

\address{$^a$ Physics Department, Stanford
          University, Stanford, CA 94305-4060, USA\\
         $^b$ Department of Physics, Brown University,
              Providence, RI 02912, USA}

\begin{abstract}
If the scale of quantum gravity is near a TeV, the LHC will be
producing one black hole (BH) about every second. The BH decays
into prompt, hard photons and charged leptons is a clean signature
with low background. The absence of significant missing energy
allows the reconstruction of the mass of the decaying BH. The
correlation between the BH mass and its temperature, deduced from
the energy spectrum of the decay products, can test
experimentally the higher dimensional Hawking evaporation law. It
can also determine the number of large new dimensions and the
scale of quantum gravity.
\end{abstract}

\maketitle

{\bf Introduction:} An exciting consequence of TeV-scale quantum
gravity \cite{add} is the possibility of production of black holes
(BHs)~\cite{adm,bf,ehm} at the LHC and beyond. The objective of
this paper is to point out the experimental signatures of BH
production. Black holes are well understood general-relativistic
objects when their mass \mbh far exceeds the fundamental (higher
dimensional) Planck mass $\MP \sim$TeV. As \mbh approaches \MP,
the BHs become ``stringy'' and their properties complex. This
raises an obstacle to calculating the production and decay of
light BHs, those most directly accessible to the LHC, where the
center-of-mass (c.o.m.) energy of colliding beams is comparable 
to the Planck mass. In what follows, we will ignore this obstacle 
and estimate the properties of light BHs by simple semiclassical 
arguments, strictly valid for $\mbh \gg \MP$. We expect that this 
will be an adequate approximation, since the important experimental
signatures rely on two simple qualitative properties: (i) the
absence of small couplings and (ii) the ``democratic" (flavor
independent) nature of BH decays, both of which may survive as
average properties of the light descendants of black holes.
Nevertheless, because of the unknown stringy corrections, our
results are approximate estimates. For this reason, we will not
attempt selective partial improvements -- such as time dependence,
angular momentum, charge, hair, and other higher-order general
relativistic refinements -- which, for light BHs, may be masked by
larger unknown stringy effects. We will focus on the production
and sudden decay of Schwarzschild black holes.

{\bf Production:} The Schwarzschild radius $R_S$ of an
$(4+n)$-dimensional black hole is given by \cite{mp}:
\vspace*{-0.1in}
\begin{equation}
    R_S = \frac{1}{\sqrt{\pi}\MP}
    \left[
      \frac{\mbh}{\MP} 
      \left( 
        \frac{8\Gamma\left(\frac{n+3}{2}\right)}{n+2}
      \right)
    \right]^\frac{1}{n+1},
\label{eq:RS}
\vspace*{-0.1in}
\end{equation}
assuming that extra dimensions are large ($\gg$ $R_S$).

\begin{figure}[tbp]
\begin{center}
\epsfxsize=3.3in
\epsffile{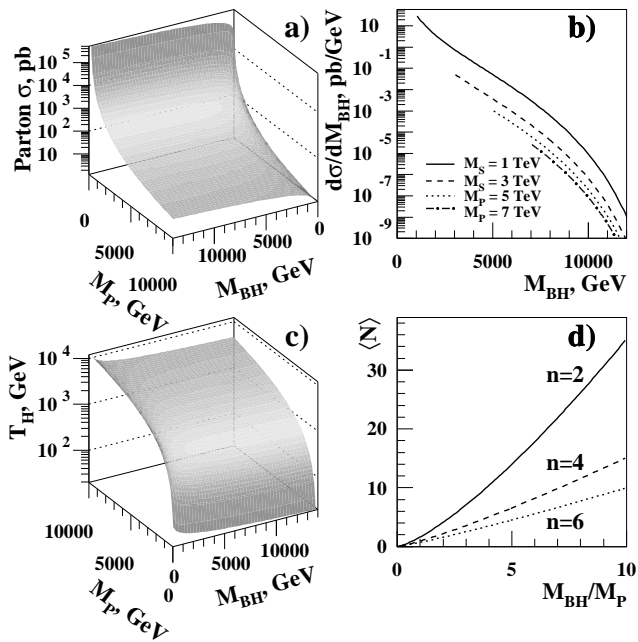}
\vspace*{-0.1in}
\caption{a) Parton-level production cross section, b) differential
cross section $d\sigma/d\MBH$ at the LHC, c) Hawking 
temperature, and d) average decay multiplicity for a Schwarzschild 
black hole. The number of extra spatial dimensions $n=4$ is used for 
a)-c). The dependence of the cross section and Hawking temperature 
on $n$ is weak and would be hardly noticable on the logarithmic scale.}
\vspace*{-0.3in}
\label{fig1}
\end{center}
\end{figure}

Consider two partons with the c.o.m. energy $\sqrt{\hat s} =
\MBH$ moving in opposite directions. Semiclassical reasoning
suggests that if the impact parameter is less than the (higher
dimensional) Schwarzschild radius, a BH with the mass \mbh forms.
Therefore the total cross section can be estimated from
geometrical arguments~\cite{footnote1}, and is of order
\vspace*{-0.1in}
\begin{equation}
    \sigma(\MBH) \approx \pi R_S^2 = \frac{1}{M_P^2}
    \left[
      \frac{\MBH}{\MP} 
      \left( 
        \frac{8\Gamma\left(\frac{n+3}{2}\right)}{n+2}
      \right)
    \right]^\frac{2}{n+1}
\label{eq:sigma}
\vspace*{-0.1in}
\end{equation}
(see Fig.~\ref{fig1}a).

This expression contains no small coupling constants; if the
parton c.o.m. energy $\sqrt{\hat s}$ reaches the fundamental
Planck scale $\MP \sim$~TeV then the cross section if of order
TeV$^{-2} \approx 400$~pb. At the LHC, with the total c.o.m. energy 
$\sqrt{s}=14$~TeV, BHs will be produced copiously. To calculate total
production cross section, we need to take into account that only a
fraction of the total c.o.m. energy in a $pp$ collision is
achieved in a parton-parton scattering. We compute the full
particle level cross section using the parton luminosity
approach (after Ref.~\cite{Barger}):
\vspace*{-0.2in}
$$
    \frac{d\sigma(pp \to \mbox{BH} + X)}{d\MBH} = 
    \frac{dL}{dM_{\rm BH}} \hat{\sigma}(ab \to \mbox{BH})
    \left|_{\hat{s}=M^2_{\rm BH}}\right.,
    \vspace*{-0.1in}
$$
where the parton luminosity $dL/d\MBH$ is defined as the sum over
all the initial parton types:
\vspace*{-0.1in}
$$
    \frac{dL}{dM_{\rm BH}} = \frac{2\MBH}{s} 
    \sum_{a,b} \int_{M^2_{\rm BH}/s}^1  
    \frac{dx_a}{x_a} f_a(x_a) f_b(\frac{M^2_{\rm BH}}{s x_a}),
    \vspace*{-0.1in}
$$
and $f_i(x_i)$ are the parton distribution functions (PDFs). We
used the MRSD$-'$~\cite{MRSD} PDF set with the $Q^2$ scale taken
to be equal to \MBH, which is within the allowed range for this
PDF set, up to the LHC kinematic limit. The dependence of the cross
section on the choice of PDF is $\sim 10\%$, i.e. satisfactory
for the purpose of this estimate.

The differential cross section $d\sigma/d\MBH$ for the BH produced
at the LHC is shown in Fig.~\ref{fig1}b for several choices of \MP.
The total production cross section at the LHC for BH masses
above \mp ranges from 0.5~nb for $\MP = 2$~TeV, $n=7$ to 120~fb for
$\MP = 6$~TeV and $n=3$. If the fundamental Planck scale is
$\approx 1$~TeV, LHC, with the peak luminosity of 30~\ifb/year
will produce over $10^7$ black holes per year. This is comparable to
the total number of $Z$'s produced at LEP, and suggests that we may
do high precision studies of TeV BH physics, as long as the backgrounds
are kept small.

{\bf Decay:} The decay of the BH is governed by its Hawking
temperature $T_H$, which is proportional to the inverse radius, and
given by~\cite{mp}:
\vspace*{-0.1in}
\begin{equation}
    T_H = \MP
    \left(
      \frac{\MP}{\MBH}\frac{n+2}{8\Gamma\left(\frac{n+3}{2}\right)}
    \right)^\frac{1}{n+1}\frac{n+1}{4\sqrt{\pi}}
\label{TH}
\vspace*{-0.1in}
\end{equation}
(see Fig.~\ref{fig1}b). As the parton collision energy increases, 
the resulting black hole gets heavier and its decay products get colder.

Note that the wavelength $\lambda = {2 \pi \over T_H}$
corresponding to the Hawking temperature is larger than the size
of the black hole. Therefore, the BH is, to first approximation, a
point-radiator and therefore emits mostly $s$-waves. This
indicates that it decays equally to a particle on the brane and in
the bulk, since it is only sensitive to the radial coordinate and
does not make use of the extra angular modes available in the
bulk. Since there are many more particles on our brane than in the
bulk, this has the crucial consequence that the black hole decays
visibly to standard model (SM) particles~\cite{ehm,lenny}.

The average multiplicity of particles produced in the process of
BH evaporation is given by: $\langle N \rangle = \left\langle
\frac{\MBH}{E} \right\rangle$, where $E$ is the energy spectrum
of the decay products. In order to find $\langle N \rangle$, we note
that the BH evaporation is a blackbody radiation process, with the 
energy flux per unit of time given by Planck's formula:
$\frac{df}{dx} \sim \frac{x^3}{e^x \pm c}$, where $x \equiv E/T_H$, and
$c$ is a constant, which depends on the quantum statistics of the
decay products ($c = -1$ for bosons, $+$1 for fermions, and 0 for
Boltzmann statistics).

The spectrum of the BH decay products in the massless
particle approximation is given by: $\frac{dN}{dE} \sim
\frac{1}{E}\frac{df}{dE} \sim \frac{x^2}{e^x \pm c}$. In order to
calculate the average multiplicity of the particles produced in
the BH decay, we use the average of the distribution in the
inverse particle energy:
\vspace*{-0.1in}
\begin{equation}
    \left\langle \frac{1}{E} \right\rangle =
    \frac{1}{T_H}\frac{\int_0^\infty dx \frac{1}{x}
    \frac{x^2}{e^x\pm c}}{\int_0^\infty dx\frac{x^2}{e^x \pm c}}
    = a/T_H,
\label{eav}
\vspace*{-0.1in}
\end{equation}
where $a$ is a dimensionless constant that depends on the type of
produced particles and numerically equals 0.68 for bosons, 0.46
for fermions, and $\frac{1}{2}$ for Boltzmann statistics. Since a
mixture of fermions and bosons is produced in the BH decay, we can
approximate the average by using Boltzmann statistics, which gives
the following formula for the average multiplicity: $\langle N
\rangle \approx \frac{\MBH}{2T_H}$. Using Eq. (\ref{TH}) for
Hawking temperature, we obtain:
\vspace*{-0.1in}
\begin{equation}
    \langle N \rangle = \frac{2\sqrt{\pi}}{n+1}
    \left(\frac{\MBH}{\MP}\right)^\frac{n+2}{n+1}
    \left(\frac{8\Gamma\left(\frac{n+3}{2}\right)}{n+2}\right)^\frac{1}{n+1}.
\label{nav}
\vspace*{-0.1in}
\end{equation}

Eq. (\ref{nav}) is reliable when the mass of the BH is much larger
than the Hawking temperature, i.e. $\langle N \rangle \gg 1$;
otherwise, the Planck spectrum is truncated at $E \approx \MBH/2$ 
by the decay kinematics~\cite{footnote2}. The average number of 
particles produced in the process of BH evaporation is shown in 
Fig.~\ref{fig1}d, as a function of $\MBH/\MP$, for several values 
of $n$.

We emphasize that, throughout this paper, we ignore time
evolution: as the BH decays, it gets lighter and hotter and its
decay accelerates. We adopt the ``sudden approximation'' in which
the BH decays, at its original temperature, into its decay
products. This approximation should be reliable as the BH
spends most of its time near its original mass and temperature,
because that is when it evolves the slowest; furthermore, that is
also when it emits the most particles. Later, when we test the
Hawking mass-temperature relation by reconstructing Wien's
dispacement law, we will minimize the sensitivity to the late and
hot stages of the BHs life by looking at only the soft
part of the decay spectrum. Proper treatment of time evolution,
for $\mbh \approx \MP$, is difficult, since it immediately takes us
to the stringy regime.

{\bf Branching Fractions:} The decay of a BH is thermal: it
obeys all local conservation laws, but otherwise does not discriminate
between particle species (of the same mass and spin). Theories with
quantum gravity near a TeV must have additional symmetries, beyond the
standard $SU(3) \times SU(2) \times U(1)$, to guarantee proton longevity,
approximate lepton number(s) and flavor conservation~\cite{footnote3}. 
There are many possibilities: discrete or continuous symmetries, four 
dimensional or higher dimensional ``bulk'' symmetries \cite{ad}. 
Each of these possible symmetries constrains the decays of the 
black holes. Since the typical decay involves a large number of  
particles, we will ignore the constraints imposed by the few conservation 
laws and assume that the BH decays with roughly equal probability to all 
off $\approx 60$ particles of the SM. Since there are six charged leptons 
and one photon, we expect $\sim 10\%$  of the particles to be hard, 
primary leptons and $\sim 2\%$ of the particles to be hard photons, each 
carrying hundreds of GeV of energy. This is a very clean signal, with 
negligible background, as the production of SM leptons or photons in 
high-multiplicity events at the LHC occurs at a much smaller rate than 
the BH production (see Fig.~\ref{nbh}). These events are also easy to 
trigger on, since they contain at least one prompt lepton or photon with 
the energy above 100 GeV, as well as energetic jets.

{\bf Test of the Hawking's radiation:} Furthermore, since there
are three neutrinos, we expect only $\sim 5\%$ average missing
transverse energy (\MET) per event, which allows us to precisely
estimate the BH mass from the visible decay products. We can also
reconstruct the BH temperature by fitting the energy spectrum of
the decay products to the Planck's formula. Simultaneous knowledge
of the BH mass and its temperature allows for a test of the Hawking's
radiation and can provide an evidence that the observed events come 
from the production of BH, and not from some other new physics.

There are a few important experimental techniques that we will use
to carry out the numerical test. First of all, to improve
precision of the BH mass reconstruction we will use only the
events with \met consistent with zero. Given the small probability
for a BH to emit a neutrino or a graviton, total statistics won't
suffer appreciably from this requirement. Since BH decays have
large jet activity, the \mbh resolution will be dominated by the
jet energy resolution and the initial state radiation effects, and
is expected to be $\sim 100$~GeV for a massive BH. Second, we will
use only photons and electrons in the final state to reconstruct the
Hawking temperature. The reason is twofold: final states with
energetic electrons and photons have very low background at high
$\sqrt{\hat s}$, and the energy resolution for electrons and
photons remains excellent even at the highest energies achieved in
the process of BH evaparation. We do not use muons, as
their momenta are determined by the track curvature in
the magnetic field, and thus the resolution deteriorates fast 
with the muon momentum growth. We also ignore the $\tau$-lepton decay 
modes, as the final states with $\tau$'s have much higher background than inclusive
electron or photon final states, and also because their energies can
not be reconstructed as well as those for the electromagnetic objects.
Fraction of electrons and photons among the final state particles
is only $\sim 5\%$, but the vast amount of BHs produced at the LHC
allows us to sacrifice the rest of the statistics to allow for a
high-precision measurement. (Also, the large number of decay particles 
enhances the probability to have a photon or an electron in the event.) 
Finally, if the energy of a decay particle approaches the kinematic 
limit for pair production, $\MBH/2$, the shape of the energy spectrum 
depends on the details of the BH decay model. In order to eliminate 
this unwanted model dependence, we use only the low part of the energy 
spectrum with $E < \MBH/2$.

The experimental procedure is straightforward: we select the BH sample
by requiring events with high mass ($> 1$ TeV) and mutiplicity of the
final state ($N \ge 4$), which contain electrons or photons with energy
$>100$~GeV. We smear the energies of the decay products with the 
resolutions typical of the LHC detectors. We bin the events in the invariant 
mass with the bin size (500 GeV) much wider than the mass resolution. 
The mass spectrum of the BHs produced at the LHC with 100~\ifb\
of integrated luminosity is shown in Fig.~\ref{nbh} for several values 
of \mp and $n$. Backgrounds from the SM $Z(ee)+$ jets and 
$\gamma +$ jets production, as estimated with PYTHIA~\cite{PYTHIA}, 
are small (see figure).

\begin{figure}[tbp]
\begin{center}
\epsfxsize=3.3in
\epsffile{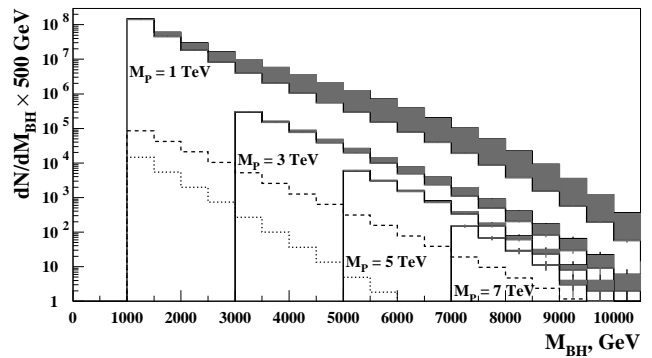}
\vspace*{-0.1in}
\caption{Number of BHs produced at the LHC in the electron or photon decay 
channels, with 100~\protect\ifb of integrated luminosity, as a function of the BH 
mass. The shaded regions correspond to the variation in the number of events 
for $n$ between 2 and 7. The dashed line shows total SM background 
(from inclusive $Z(ee)$ and direct photon production). The dotted line 
corresponds  to the $Z(ee)+X$ background alone.}
\label{nbh}
\vspace*{-0.3in}
\end{center}
\end{figure}

To determine the Hawking temperature in each \mbh bin, we perform a maximum 
likelihood fit of the energy spectrum of electrons and photons in the BH 
events to the Planck formula (with the coefficient $c$ determined by the
particle spin), below the kinematic cutoff (\MBH/2). We then
use the measured \mbh vs. \TH\ dependence and Eq.~(\ref{TH}) to determine
the fundamental Planck scale \mp and the dimensionality of space $n$.
Note that to determine $n$ we can also take the logarithm of both
sides of Eq.~(\ref{TH}):
\vspace*{-0.05in}
\begin{equation}
  \log(T_H) = \frac{-1}{n+1}\log(\MBH) + \mbox{const},
\label{logTH}
\vspace*{-0.05in}
\end{equation}
where the constant does not depend on the BH mass, but only
on \mp and on detailed properties of the bulk space, such as shape of
extra dimensions. Therefore, the slope of a straight-line fit to
the $\log(\TH)$ vs. $\log(\MBH)$ data offers a direct way of determining
the dimensionality of space. This is a multidimensional analog of Wien's
displacement law. Note that Eq.~(\ref{logTH}) is fundamentally different
from other ways of determining the dimensionality of space-time, e.g.
by studying a monojet signature or a virtual graviton exchange processes,
also predicted by theories with large extra dimensions.

\begin{figure}[tbp]
\begin{center}
\epsfxsize=3.3in
\epsffile{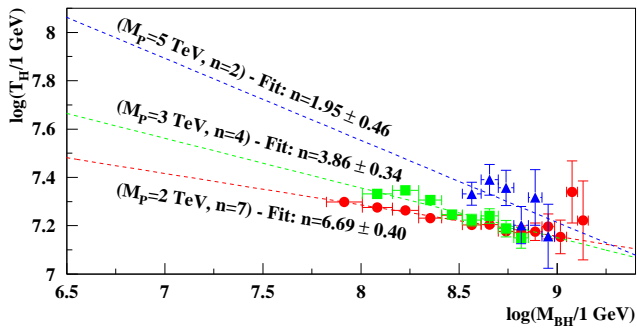}
\caption{Determination of the
dimensionality of  space via Wien's displacement law at the LHC with
100~\protect\ifb\ of data.}
\label{Wien}
\vspace*{-0.3in}
\end{center}
\end{figure}

Test of the Wien's law at the LHC would provide a confirmation
that the observed $e+X$ and $\gamma+X$ event excess is due to the
BH production. It would also be the first experimental test of the
Hawking's radiation hypothesis. Figure~\ref{Wien} shows typical fits
to the simulated BH data at the LHC, corresponding to 100~\ifb\ of
integrated luminosity, for the highest fundamental Planck scales
that still allow for determination of the dimensionality of space
with reasonable precision. The reach of the LHC for the
fundamental Planck scale and the number of extra dimensions via
Hawking's radiation extends to $\MP \sim 5$~TeV and
is summarized in Table~\ref{table}~\cite{footnote4}.

\begin{table}[tb]
\begin{center}
\caption{Determination of \protect\mp and $n$ from Hawking's radiation. 
The two numbers in each column correspond to fractional 
uncertainty in \protect\mp and absolute uncertainty in $n$, respectively.}
\begin{tabular}{cccccc}
\hline
~~\MP       &	  1 TeV  &  2 TeV   & 3 TeV      &  4 TeV    & 5 TeV      \\
\hline
$n=2$     & 1\%/0.01 &  1\%/0.02& 3.3\%/0.10 & 16\%/0.35 & 40\%/0.46  \\
$n=3$     & 1\%/0.01 & 1.4\%/0.06 & 7.5\%/0.22 & 30\%/1.0 & 48\%/1.2  \\
$n=4$     & 1\%/0.01 & 2.3\%/0.13 & 9.5\%/0.34 & 35\%/1.5 & 54\%/2.0  \\
$n=5$     & 1\%/0.02 & 3.2\%/0.23 & 17\%/1.1 \\
$n=6$     & 1\%/0.03 & 4.2\%/0.34 & 23\%/2.5  & \multicolumn{2}{c}{Fit fails}\\
$n=7$     & 1\%/0.07 & 4.5\%/0.40 & 24\%/3.8 \\
\hline
\end{tabular}
\end{center}
\label{table}
\vspace*{-0.3in}
\end{table}

Note, that the BH discovery potential at the LHC is maximized in the
$e/\mu+X$ channels, where background is much smaller than that 
in the $\gamma+X$ channel (see Fig.~\ref{nbh}). The reach of a simple 
counting experiment extends up to $\MP \approx 9$ TeV ($n=2$--7),
where one would expect to see a handful of BH events with negligible 
background.

{\bf Summary:} Black hole production at the LHC may be one of the early
signatures of TeV-scale quantum gravity. It has three advantages:

{\bf Large Cross Section.} Because no small dimensionless coupling constants, analogous to $\alpha$, suppress the production of BHs. This leads to enormous rates.

{\bf Hard, Prompt, Charged Leptons and Photons.} Because thermal decays are flavor-blind. This signature has practically vanishing SM background.

{\bf Little Missing Energy.} This facilitates the determination of the mass and the temperature of the black hole, and may lead to a test of Hawking's radiation.

It is desirable to improve our primitive estimates, especially for the
light black holes ($\MBH \sim \MP$); this will involve string theory.
Nevertheless, the most telling signatures of BH production~-- large and
growing cross sections; hard leptons, photons, and jets~-- emerge from
qualitative features that are expected to be reliably estimated from the
semiclassical arguments of this paper.

Perhaps black holes will be the first signal of TeV-scale quantum gravity. This
depends on, among other factors, the relative magnitude of $M_P$ and the
(smaller) string scale $M_S$. For $M_S \ll \MP$, the vibrational modes of 
the string may be the first indication of the new physics.

Note added: After the completion of this work, a related paper~\cite{gt}
has appeared in the LANL archives.

{\bf Acknowledgments:} We would like to thank Gia Dvali, Veronika Hubeny,
Nemanja Kaloper, Elias Kiritsis, Konstantin Matchev, and Lenny Susskind for
valuable conversations, and many of the participants of the ``Physics at TeV 
Colliders'' and ``Avatars of M-Theory'' workshops for their interest.


\newpage
\begin{thebibliography}{15}
\bibitem{add}
N.~Arkani-Hamed, S.~Dimopoulos, and G.~Dvali, Phys. Lett.
B {\bf 429}, 263 (1998); I.~Antoniadis, N.~Arkani-Hamed,
S.~Dimopoulos, and G.~Dvali, Phys. Lett. B {\bf 436}, 257
(1998); N.~Arkani-Hamed, S.~Dimopoulos, and G.~Dvali,
Phys. Rev. D {\bf 59}, 086004 (1999).

\bibitem{adm}
P.C.~Argyres, S.~Dimopoulos, and J.~March-Russell, Phys. Lett. {\bf B441}, 96 (1998).

\bibitem{bf}
T.~Banks and W.~Fischler,
hep-th/9906038.

\bibitem{ehm}
R.~Emparan, G.T.~Horowitz, and R.C.~Myers,
Phys. Rev. Lett. {\bf 85}, 499 (2000) [hep-th/0003118].

\bibitem{mp}
R.C. Myers and M.J. Perry, {\it Ann. Phys.}\/ {\bf 172}, 304 (1986).

\bibitem{footnote1}
In fact the cross section is somewhat enhanced by initial-state attraction 
\protect\cite{ehm}.

\bibitem{Barger}
V.D.~Barger and R.J.N.~Phillips, ``Collider Physics, Updated Edition,'' Addison-Wesley, 1997, p.~157.

\bibitem{MRSD}
A.D.~Martin, R.G.~Roberts, and W.J.~Stirling, preprint RAL-92-078 (1992).

\bibitem{lenny} L. Susskind, private communication.

\bibitem{footnote2}
To avoid this limitation when performing numerical tests of Wien's law, 
we truncate the integrals in Eq. (\protect\ref{eav}) at the kinematic 
limit (\MBH/2).

\bibitem{footnote3}
Any form of new physics at a TeV requires such new symmetries. 
An example, for TeV-scale supersymmetry, is R-parity.

\bibitem{ad}
N.~Arkani-Hamed and S.~Dimopoulos, hep-ph/9811353.

\bibitem{PYTHIA}
T.~Sj\"{o}strand, Comp.~Phys.~Comm. {\bf 82}, 74 (1994).

\bibitem{footnote4}
We checked the effect of introducing  an additional cutoff at $E < M_S$, 
where string dynamics is expected to affect the shape of the Planck
spectrum,  by taking $M_S =\MP/2$. As a result, the uncertainties in 
\mp and $n$ increased by about a factor of 2.

\bibitem{gt}
S.B.~Giddings and S.~Thomas, hep-ph/0106219, v2.
\end{thebibliography}
\end{document}